\documentclass{ws-procs9x6}
\begin{document}

\title{Some recent results in calculation of the Casimir energy at
zero and finite temperature}

\author{V.~V. NETERENKO\footnote{\uppercase{W}ork partially
supported by grant 03-01-00025 of the \uppercase{R}ussian
\uppercase{F}oundation for \uppercase{B}asic \uppercase{R}search and by
\uppercase{ISTC} (\uppercase{P}roject  840).}}

\address{Bogoliubov Laboratory for Theoretical Physics,\\ Joint Institute
for Nuclear Research, \\
Dubna, 141980, Russia\\
E-mail: nestr@thsun1.jinr.ru}


\maketitle

\abstracts{The survey summarizes briefly the results obtained recently in the
 Casimir effect studies considering the following subjects:
   i) account of the material characteristics of the media and
their influence on the vacuum energy (for example, dilute
dielectric ball);
   ii) application of the spectral geometry methods for investigating
the vacuum energy of quantized fields with the goal to gain some
insight, specifically, in the geometrical origin of the
divergences that enter the vacuum energy and to develop the
relevant renormalization procedure;
  iii) universal method for calculating the high temperature dependence
of the Casimir energy  in terms of heat kernel coefficients.}

\section{Casimir energy of a dilute dielectric ball}
Calculation of the Casimir energy of a dielectric ball has a rather long
history starting 20 years ago.\cite{1} However only recently the final result
was obtained for a dilute dielectric ball at zero\cite{2,LSN} and
finite\cite{NLS,Barton} temperature. Here we summarize briefly the derivation
of the Casimir energy of a dilute dielectric ball by making use of the mode
summation method and the addition theorem for the Bessel functions instead of
the uniform asymptotic expansion for these functions.\cite{LSN,NLS}

A  solid ball of radius $a$ placed in an unbounded uniform medium is
considered. The contour integration technique\cite{LSN}  gives ultimately the
following representation for the Casimir energy of the ball
\begin{equation}
 \label{1}
  E=-\frac{1}{2\pi a}\sum_{l=1}^{\infty}(2l+1)\int_0^{y_0}\!\!\!dy\,
 y\,\frac{d}{dy}\ln\left[W_l^2(n_1y, n_2y)-\frac{\Delta n^2}{4}
 P_l^2(n_1y, n_2y)\right],
\end{equation}
where
\begin{eqnarray}
  W_l(n_1y,
n_2y)&=&s_l(n_1y)e_l^{\prime}(n_2y)-s_l^{\prime}(n_1y)e_l(n_2y)\,,
  \nonumber \\
  P_l(n_1y,
n_2y)&=&s_l(n_1y)e_l^{\prime}(n_2y)+s_l^{\prime}(n_1y)e_l(n_2y)\,,
\nonumber
\end{eqnarray}
and $s_l(x)$, $e_l(x)$ are the modified Riccati-Bessel functions,
$n_1, n_2 $ are the refractive indices of the ball and of its
surroundings, $\Delta n= n_1-n_2$.

Analysis of divergences\cite{LSN} leads to the following algorithm for
calculating the vacuum energy (\ref{1}) in the $\Delta n^2$-approximation.
First, the $\Delta n^2$--contribution should be found, which is given by the
sum $\sum_lW_l^2$. Upon changing its sign to the opposite one, we obtain the
contribution generated by $W_l^2$, when this function is in the argument of the
logarithm. The $P^2_l$-contribution into the vacuum energy is taken into
account by expansion of Eq. (\ref{1}) in terms of $\Delta n^2$.

 Applying the addition theorem for the Bessel
functions
\[
  \sum_{l=0}^{\infty}(2l+1)[s_l^{\prime}(\lambda
  r)e_l(\lambda\rho)]^2=\frac{1}{2r \rho}\int_{r-\rho}^{r+\rho}
  \left(\frac{1}{\lambda}\,\frac{\partial{G}}{\partial r}
 \right)^2R\;dR
\]
with
\[
{G} ={\lambda r\rho}{R^{-1}}\, e^{-\lambda R}, \quad
  R=\sqrt{r^2+\rho^2-2r\rho\cos\theta}
\]
one arrives at the result
\[
  E=\frac{23}{384}\frac{\Delta n^2}{\pi a}
  =\frac{23}{1536}\,\frac{(\varepsilon_1-
  \varepsilon_2)^2}{\pi a}{,} \quad  \varepsilon_i=n^2_i, \quad i=1,2\,{.}
\]

Extension to finite temperature $T$ is accomplished by
substituting the $y$-integra\-tion in (\ref{1}) by summation over
the Matsubara frequencies $ \omega_n=2\pi nT$.

In the $\Delta^2$-approximation the last term in Eq.\ (3.20) from the
article\cite{NLS}
\begin{equation}
\label{eq-2} \overline {U}_W(T)=2 T\Delta n^2 \sum_{n=0}^\infty
\!{}^{'}w^2_n\int_{\Delta n}^2\frac{e^{-2w_nR}}{R}\,dR{,} \quad
w_n=2 \pi na T
\end{equation}
can be represented in the following form
\begin{equation}
\label{eq-3} \overline {U}_W(T)=-2 T\Delta n^2
\sum_{n=0}^\infty\!{}^{'} w^2_n \,E_1(4w_n){,}
\end{equation}
where $E_1(x)$ is the exponential-integral function. Now we accomplish the
summation over the Matsubara frequencies by making use of the Abel-Plana
formula
\begin{equation}
\label{eq-4} \sum_{n=0}^\infty\!{}^{'} f(n) =\int_0^\infty
f(x)\,dx+i\int_0^\infty \frac{f(ix)-f(-ix)}{e^{2\pi x}-1}\;dx{.}
\end{equation}
The first term in the right-hand side of this equation gives the
contribution independent of the temperature, and the net
temperature dependence is produced by the second term in this
formula. Being interested in the low temperature behavior of the
internal energy we substitute into the second term in Eq.\
(\ref{eq-4}) the following  expansion of the function $E_1(z)$
\begin{equation}
\label{eq-5} E_1(z) =-\gamma -\ln z- \sum^\infty_{k=1}\frac{(-1)^k
z^k}{k\cdot k!},\quad |\arg z |<\pi {,}
\end{equation}
where $\gamma $ is the Euler constant. The contribution proportional to $T^3$
is produced by the logarithmic term in the expansion (\ref{eq-5}). The higher
powers of $T$ are generated by the respective terms in the sum over $k$ in this
formula $(t=2\pi a T)$
\begin{equation}
\label{eq-6} \overline {U}_W(T)=\frac{\Delta n^2}{\pi a}\left ( -\frac{1}{96}+
\frac{\zeta (3)}{4\pi ^2} t^3 -\frac{1}{30}t^4 +\frac{8}{567} t^6
-\frac{8}{1125}t^8+{O}(t^{10})\right ) {.}
\end{equation}

Taking all this into account we arrive at the following low
temperature behavior of the internal Casimir energy of a dilute
dielectric  ball
\begin{equation}
\label{eq-7} U(T)= \frac{\Delta n^2}{\pi a}\left ( \frac{23}{384}
+\frac{\zeta(3)}{4\pi^2}t^3 -\frac{7}{360}t^4 +\frac{22}{2835}t^6
-\frac{46}{7875}t^8 +{O}(t^{10})
 \right ){.}
\end{equation}
The relevant  thermodynamic relation give the following low temperature
expansions for free energy
\begin{equation}
\label{eq-8} F(T)=\frac{\Delta n^2}{\pi a}\left (
\frac{23}{384}-\frac{\zeta (3)}{8\pi ^2}t^3+\frac{7}{1080}t^4
 -\frac{22}{14175}t^6+{O}(t^{8})
\right )\,{.}
\end{equation}

For large temperature $T$ we found\cite{NLS}
\begin{equation}
\label{eq-10}
 U(T) \simeq  {\Delta n^2}T/8 {,}\;\;
 F(T)  \simeq  -{\Delta n^2} T\left [\ln (aT)-c\right ]/8{,}
\end{equation}
where $c$ is a constant\cite{Barton,BNP} $ c=\ln 4 +\gamma -{7}/{8}\,{.}$
Analysis of Eqs.\ (3.20) and (3.31) from the paper\cite{Barton} shows that
there are only exponentially suppressed corrections to the leading terms
(\ref{eq-10}).

Summarizing we conclude that now there is a complete agreement between the
results of  calculation of  the Casimir thermodynamic functions for a dilute
dielectric ball carried out in the framework of two different approaches:  by
the mode summation method\cite{LSN,NLS} and by perturbation theory for
quantized electromagnetic field, when  dielectric ball is considered as a
perturbation in unbounded continuous surroundings.\cite{Barton}
\section{Spectral geometry and vacuum energy}
\label{sec:333}
 In spite of a quite long history of the Casimir
effect (more than 50 years) deep understanding and physical intuition in this
field are still lacking. The main problem here is the separation of net finite
effect from the divergences inevitably present in the Casimir calculations. A
convenient analysis of these divergences is provided by the heat kernel
technique, namely, the coefficients of the asymptotic expansion of the heat
kernel.

Keeping in mind the elucidation of the origin of these divergences in
paper\cite{s-c} the vacuum energy of electromagnetic field has been calculated
for a semi-circular infinite cylindrical shell. This shell is obtained by
crossing an infinite cylinder by a plane passing through  its symmetry axes. In
the theory of waveguides it is well known that a semi-circular waveguide has
the same eigenfrequencies as the cylindrical one but without degeneracy
(without doubling) and safe for one frequency series. Notwithstanding the very
close spectra, the vacuum divergences in these problems prove to be drastically
different, so the zeta function technique does not give a finite result for a
semi-circular cylinder unlike for a circular one.

 It was revealed that the origin of these divergences is
the corners in the boundary of semi-circular cylinder.\cite{NPD} In terms of
the heat kernel coefficients, it implies that the coefficient $a_2$ for a
semi-circular cylinder does not vanish due to these corners.

However in the 2-dimensional (plane) version of these problems the
origin of nonvanishing $a_2$ coefficient for a semicircle  is the
contribution due to the curvature of the boundary, while the
corner contributions  to $a_2$ in 2 dimensions are cancelled.

Different geometrical origins of the vacuum divergences in the
two- and three-dimensional versions of the boundary value problem
in question evidently imply  the impossibility of obtaining a
finite and unique value of the Casimir energy  by taking advantage
of the atomic structure of the boundary or its quantum
fluctuations. It is clear, because any physical reason of the
divergences should hold simultaneously in the two- and
three-dimensional versions of a given boundary configuration.
\section{High temperature asymptotics
of vacuum energy in terms of  heat kernel coefficients}
\label{sec:444} The Casimir calculations at finite temperature
prove to be a nontrivial problem specifically for boundary
conditions with nonzero curvature.  For this goal a powerful
method of the zeta function technique and the heat kernel
expansion can be used. For obtaining the high temperature
asymptotics of the thermodynamic characteristics it is sufficient
to know the heat kernel coefficients and the determinant   for the
spatial part of the operator governing the field dynamics.
 This is an essential merit
of this approach.\cite{BNP}

Starting point is the general high temperature expansion of the free energy in
terms of the heat kernel coefficients.\cite{DK} These  coefficients needed for
are calculated as the residua of the corresponding zeta functions.\cite{LNB}

\subsection{A perfectly conducting spherical shell  in vacuum}
The first six heat kernel coefficients in this problem are:
\begin{eqnarray}
a_0=a_{1/2}=a_1=0, \;\; \frac{a_{3/2}}{(4\pi)^{3/2}}= \frac{1}{4}, \;\;
a_2=0,\;\; \frac{a_{5/2}}{(4\pi)^{3/2}}=\frac{c^2}{160\,R^2}. \label{eq_2}
\end{eqnarray}
Furthermore
\begin{equation}
a_j=0,\qquad j=3,4,5,  \dots\, {.} \label{eq3}
\end{equation}
 The exact value of  $\zeta' (0)$ is
derived in Ref.\cite{BNP}
\begin{equation}
\zeta' (0)= 0.38429+({1}/{2})\ln({R}/{c}). \label{eq4}
\end{equation}
As a result we have the following high temperature limit for the free energy
\begin{equation}
F (T)=-\frac{T}{4}\left({\displaystyle 0.76858}+\ln \tau+
 \frac{1}{960\tau ^2}\right )+{O}(T^{-3}),
\label{eq5}
\end{equation}
where $\tau =RT/(\hbar c)$ is the dimensionless `temperature'. The expression
(\ref{eq5}) exactly reproduces the  asymptotics obtained in Ref.\cite{BD} by
making use of the multiple scattering technique (see Eq.~(8.39) in that paper).
\subsection{A perfectly conducting cylindrical shell} The heat kernel
coefficients are
\begin{equation}
a_0= a_{1/2}= a_1= a_2=0,\quad
\frac{a_{3/2}}{(4\pi)^{3/2}}=\frac{3}{64\,R}, \quad
\frac{a_{5/2}}{(4\pi)^{3/2}}=\frac{153}{8192}\frac{c^2}{R^3}.
\end{equation}
The zeta function determinant in this problem is calculated in Ref.\cite{BNP}
\begin{equation}
\zeta'(0)={0.45711}/{R}+({3}/{32\,R})\,\ln({R}/{2\,c}). \label{eq5_6}
\end{equation}
The free energy behavior at high temperature is the following
\begin{equation}
F(T)=-\frac{T}{R} \left (0.22856 +\frac{3}{64} \ln\frac{\tau
}{2}-\frac{51}{65536\tau ^2}\right ) +{O}(T^{-3}). \label{fcs}
\end{equation}
 The high temperature asymptotics  of the electromagnetic
free energy in presence of perfectly conducting cylindrical shell was
investigated  in paper.\cite{BD} To make the comparison handy let us rewrite
their result as follows
\begin{equation}
F(T)\simeq-({T}/{R})\left [0.10362+({3}/{64R}) \ln({\tau}/{2})\right ].
\label{15a}
\end{equation}
The discrepancy between the  terms linear  in $T$ in Eqs.\ (\ref{fcs}) and
(\ref{15a}) is due to the double scattering approximation used in Ref.\cite{BD}
Our approach gives the exact value of this term (see Eq.\ (\ref{fcs})).

\section{Conclusion}
The inferences concerning the individual subjects of this review have been done
in respective sections. Here we only  note, that in order to cast the theory of
the Casimir effect  to a complete form further studies are certainly needed.

\end{document}